\documentclass[aps,epsf,floatfix,pre,11pt]{revtex4-2}
\usepackage{graphicx}
\usepackage{amsmath}
\usepackage{amssymb}
\usepackage{multirow}
\usepackage{xcolor}
\usepackage{hyperref}
\usepackage{float}
\usepackage{mathtools}

\usepackage{bm}
\usepackage{matlab-prettifier}
\usepackage{soul}
\allowdisplaybreaks
\usepackage{ulem}

\begin{document}
\title{Molecular memory reduces timing noise in stochastic gene expression}

\author{Kuheli Biswas}
\affiliation{Department of Physics of Complex Systems, Weizmann Institute of Science, Rehovot 7610001, Israel}

\author{Supravat Dey}
\affiliation{Department of Physics, SRM University-AP, Neerukonda, Andhra Pradesh 522240, India}

\author{Abhyudai Singh}
\affiliation{Departments of Electrical and Computer Engineering, Biomedical Engineering, and Mathematical Sciences, Center for Bioinformatics and Computational Biology, University of Delaware, Newark, DE 19716, USA}

\begin{abstract}
Fluctuations in gene expression affect the timing precision of downstream events triggered when a gene product reaches a critical threshold. The first-passage time (FPT) to this threshold characterizes the timing of such events. In conventional single-step birth--death models, reaction waiting times are exponentially distributed and hence memoryless. In this setting, FPT noise exhibits a U-shaped dependence on the threshold, attaining its minimum near one-half of the steady-state mean. However, mRNA production and degradation often involve multiple sequential steps, leading to hypoexponential distributions of mRNA production times and mRNA lifetimes. Here, we investigate how molecular memory arising from multistep mRNA production and degradation affects FPT statistics in bursty gene expression. We calculate FPT statistics in the small-noise limit. Both multistep mechanisms reduce FPT noise, but through distinct mechanisms. Multistep burst arrival reduces mRNA copy-number fluctuations, thereby suppressing timing noise. Multistep degradation instead reshapes transient accumulation dynamics, reducing relative timing noise even when steady-state mRNA noise increases. The threshold that minimizes FPT noise remains close to one-half of the steady-state mean, except for multistep burst arrival with constant burst sizes, for which it shifts to lower values. Stochastic simulations support the analytical predictions within their regimes of validity and show that these qualitative conclusions persist beyond the small-noise approximation.
\end{abstract}

\maketitle
\section{Introduction}

First-passage processes provide a unifying framework for stochastic phenomena in which a fluctuating variable first reaches a prescribed boundary. Such problems arise throughout statistical physics, from diffusion-controlled reactions and target search to escape processes and neuronal firing~\cite{Redner2001FirstPassage, Benichou2011Intermittent}. The same framework is particularly natural for intracellular processes, where a cellular event is often initiated when the abundance of a regulatory molecule crosses a critical threshold \cite{IyerBiswasZilman2016, Ghusinga2017FirstPassage, Ham2024}. When the regulatory molecule is a gene product, fluctuations in its threshold-crossing time arise from the intrinsic stochastic biochemical reactions underlying gene expression~\cite{Paulsson2005Models}. The noise in the corresponding first-passage time (FPT) therefore quantifies the temporal precision of the downstream event. In simple memoryless birth--death models of gene expression, for which reaction waiting times are exponentially distributed, FPT noise depends nonmonotonically on the normalized threshold and is minimized when the threshold is close to one-half of the steady-state mean~\cite{Ghusinga2017FirstPassage, DalCo2017StochasticTiming}.
This nonmonotonic behavior has also been observed experimentally in bacteriophage $\lambda$ lysis timing by altering the threshold through sequence mutations in holin~\cite{Kannoly2020OptimumThreshold}.

The study of threshold-crossing dynamics has expanded beyond the minimal birth--death model to encompass more complex biological processes and regulatory mechanisms. FPT approaches have been applied to bacterial cell-division control and to gene-expression models incorporating promoter switching, external disturbances, and cell growth~\cite{Ghusinga2016CellDivision, GhusingaSingh2019GeneSwitching, GhusingaSingh2020ExternalDisturbances, Rijal2022ExactDistribution}. Experimental studies of bacteriophage $\lambda$ lysis have further demonstrated how the holin threshold and translational efficiency influence lysis-time statistics~\cite{Kannoly2020OptimumThreshold, Kannoly2025TranslationEfficiency}. The role of regulatory mechanisms involving feedback loops, post-transcriptional regulation, and molecular sequestration in shaping FPT statistics has also been extensively investigated~\cite{Dey2021Feedforward, Vahdat2021Feedback, biswas2020timing, Rezaee2023SelfRegulated, Rezaee2024RegulatedDegradation, Biswas2024Decoys}. Collectively, these studies have advanced our understanding of how stochastic gene-expression dynamics and cellular regulation influence the timing of intracellular events. 

Despite these advances, most theoretical treatments of FPT statistics still model molecular production and degradation as single-step, memoryless processes with exponentially distributed waiting times. Experimental observations and mechanistic models, however, support multistep descriptions of both transcription and mRNA degradation~\cite{Pedraza2008MolecularMemory, Deneke2013ComplexDegradation, Weidemann2023SubPoissonian}.
When the intermediate states are coarse-grained, the total waiting time for a sequence of exponential steps is generally hypoexponential, introducing effective molecular memory~\cite{ZhangZhou2019MolecularMemory, szavitsnossan2024solving}.
At the production level, the intermediate states may correspond to transcription initiation, pre-mRNA processing, and nuclear export; at the degradation level, they may correspond to deadenylation, decapping, and exonucleolytic decay~\cite{Bentley2014RNAProcessing, Garneau2007mRNADecay, Ma2024}. 
Queueing-theoretic approaches have related general interarrival-time and lifetime distributions to steady-state copy-number moments~\cite{jia2011intrinsic}. However, how memory in burst arrival and mRNA degradation reshapes FPT statistics remains largely unexplored.

Here, we investigate how two sources of molecular memory---multistep burst arrival and multistep mRNA degradation---reshape FPT statistics in stochastic gene expression. We derive transient moments of the mRNA abundance from the chemical master equation and use them to obtain analytical expressions for the mean and noise of the FPT in the small-noise limit. For multistep burst arrival, the analytical calculation of moment dynamics assumes that the mean interburst interval is short relative to the mRNA lifetime, a regime relevant to many mammalian genes~\cite{Suter2011BurstingKinetics, Halpern2015BurstyLiver, Agarwal2022mRNADecayMammals, Meeussen2024TimeWillTell}. At a fixed mean burst arrival time and mean mRNA lifetime, both multistep burst arrival and multistep degradation reduce FPT noise, but through distinct mechanisms. Multistep burst arrival regularizes production and lowers copy-number noise, thereby directly suppressing timing noise. Multistep degradation accelerates the mean mRNA response, increasing its logarithmic sensitivity to time at a fixed threshold and thereby allowing relative timing noise to decrease even when steady-state copy-number noise increases. For constant burst sizes, multistep arrival shifts the optimal normalized threshold below the Markovian value $X_0\simeq0.55$, whereas multistep degradation leaves the optimum close to or above one-half. Stochastic simulations support the analytical predictions within their expected regimes and show that the qualitative trends persist beyond the strict timescale-separation regime. Thus, burst-arrival and degradation memory provide distinct routes for controlling both temporal precision and the optimal triggering threshold.

\begin{table}[ht!]
\centering
\begin{tabular}{|c|c|c|c|}
\hline
Symbol
& Description
& Symbol
& Description
\\
\hline
$k$
& mRNA production rate
& $B$
& Burst size
\\
\hline
$\mu$
& mRNA degradation rate
& $N$
& \# sequential transcriptional steps
\\
\hline
$x(t)$
& Total mRNA copy number at time $t$
& $R$
& \# mRNA degradation steps
\\
\hline
$\langle x(t)\rangle$
& Mean of mRNA at $t$
& $\langle x_s\rangle$
& Mean steady-state of mRNA
\\
\hline
CV$_{x}(t)$
& Coefficient of variation of mRNA at time $t$
&CV$_{x_s}$
& Steady-state CV of mRNA
\\
\hline
$\sigma^2$
& Variance
&FF
& Fano factor
\\
\hline
$X$
& Threshold number of mRNA
&$X_0$
& Minimum normalized threshold $X_{min}/\langle x_{s}\rangle$
\\
\hline
$T$
& First passage time
&$CV_T$
& Coefficient of variation of FPT
\\
\hline
\end{tabular}
\caption{Summary of notation used}
\label{table:parameter}
\end{table}
\section{First-Passage-Time Framework}

We define the first-passage time (FPT) as the minimum time required for the mRNA copy number to reach a critical threshold for the first time. Starting from zero initial conditions, the FPT is
\begin{equation}\label{eq:def_FPT}
T := \inf\{t:x(t) \ge X|x(0) = 0\},
\end{equation}
where $X$ is the threshold level required to trigger a downstream event. Since gene expression is stochastic, $T$ varies between realizations. When copy-number fluctuations are small, the mean FPT, $\langle T\rangle$, can be approximated by the time at which the mean mRNA trajectory reaches the threshold,
\begin{eqnarray}
\langle T \rangle &\approx& {t|\langle x(t) \rangle = X}.
\label{eq:mean_FPT_def}
\end{eqnarray}

Timing noise is quantified by the squared coefficient of variation of the FPT, $CV_T^2 = \sigma_T^2/\langle T \rangle^2$, the ratio between variance and mean squared.
Using the small-noise approximation~\cite{DalCo2017} the FPT noise can be written as
\begin{eqnarray}
CV_T^2 &\approx&
\left[\left(S_{x}\right)^{-1} CV_{x}(t)\right]^2_{t=\langle T \rangle}, \qquad \text{where $S_{x} := \frac{d \log \langle x(t)\rangle}{d\log\left(t\right)}$}
\label{eq:noise_FPT_def}
\end{eqnarray}
is the logarithmic sensitivity of the mean mRNA level with respect to time. We emphasize that the formula given by Eq.~(\ref{eq:noise_FPT_def}) is valid in the limit of small fluctuations in copy number. Nevertheless, in the small-noise limit, the FPT noise admits a clear physical interpretation through its decomposition into two contributions: the first is determined by the sensitivity of the mean copy-number dynamics to time, while the second is determined by the relative copy-number fluctuations at the mean FPT. 
While more general approaches have been developed to approximate FPT distributions directly from master equations or Fokker--Planck equations~\cite{Chen2026FPT}, our analysis focuses on the relationship between FPT noise and mRNA moment dynamics. 
For the multi-state burst-arrival and multistage-degradation processes considered here, we validate the small-noise approximation directly against exact SSA simulation results.

\section{Multistep Burst Arrival Reduces Timing Noise}\label{sec:multi_step_arrival}

\begin{figure}[ht!]
\centering
\includegraphics[width=1\linewidth]{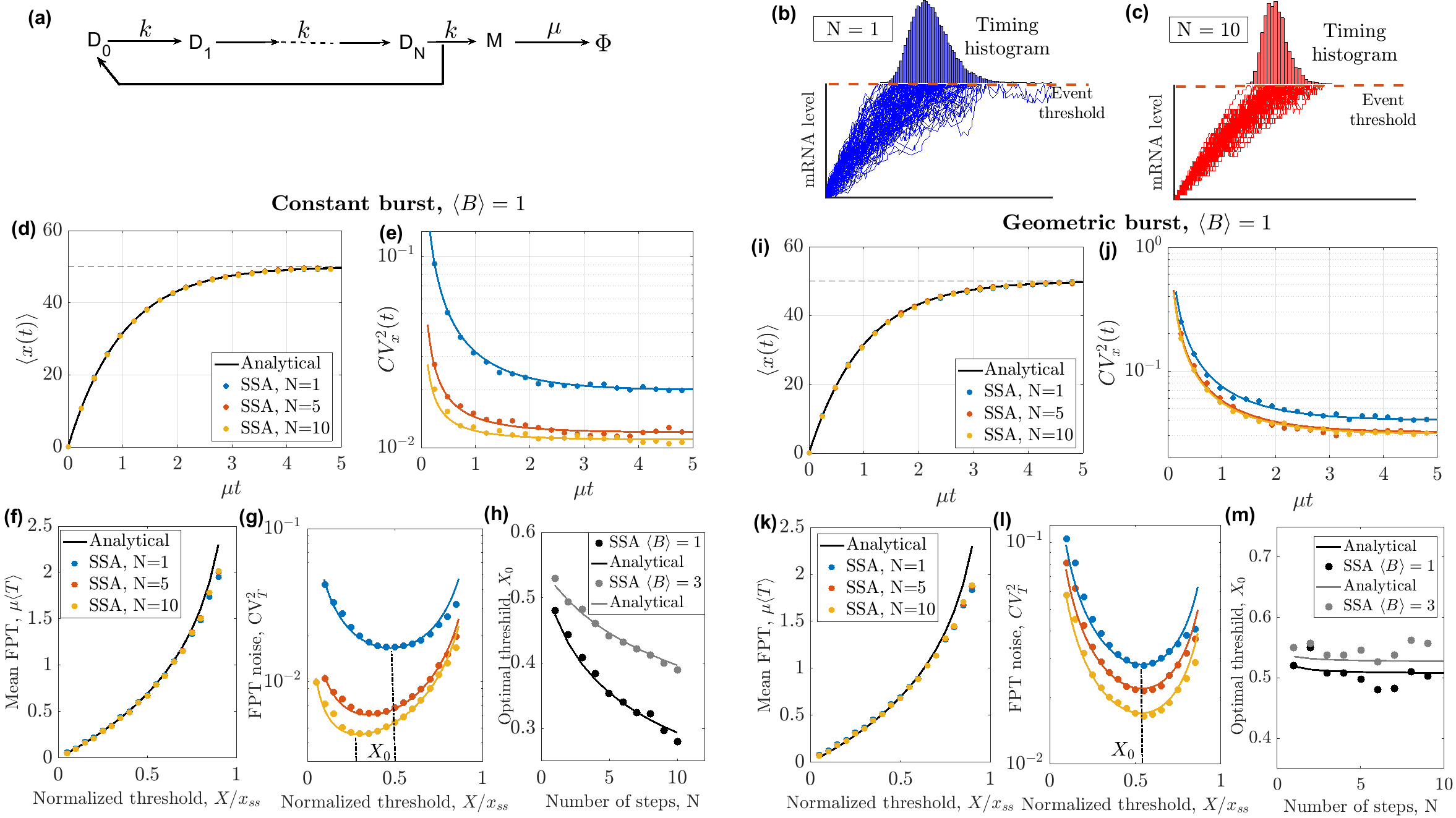}
\caption{\textbf{Multistep burst arrival suppresses both mRNA abundance noise and first-passage-time noise.} \textbf{(a)} A schematic of \(N\)-state transcription process. The nodes \(D_i\), \(i=0,\ldots,N-1\), denote successive transcriptional states, and transitions occur at rate \(k\). Completion of the final transition produces an mRNA burst of size \(B\) and resets the system to \(D_0\). Individual mRNAs undergo single-step degradation at rate \(\mu\). 
\textbf{(b,c)} Representative stochastic mRNA trajectories and corresponding first-passage-time distributions for \textbf{(b)} \(N=1\) and \textbf{(c)} \(N=10\), evaluated at the same event threshold. Increasing \(N\) reduces mRNA noise and narrows the timing distribution. 
\textbf{(d--h)} Results for constant burst. \textbf{(d)} Mean mRNA abundance and \textbf{(e)} squared coefficient of variation, \(CV_x^2(t)\), as functions of dimensionless time \(\mu t\). \textbf{(f)} Dimensionless mean FPT, \(\mu\langle T\rangle\), and \textbf{(g)} FPT noise, \(CV_T^2\), as functions of the normalized threshold \(X/\langle x_s\rangle\). \textbf{(h)} Optimal normalized threshold, \(X_0\), where FPT noise is minimum, as a function of \(N\) for \(\langle B\rangle=1\) and \(3\). 
\textbf{(i--m)} Corresponding results for zero-inclusive geometrically distributed bursts. \textbf{(i)} Mean mRNA abundance, \textbf{(j)} mRNA noise, \textbf{(k)} dimensionless mean FPT, \textbf{(l)} FPT noise, and \textbf{(m)} optimal normalized threshold. Solid curves show the analytical approximations in Eqs.~(\ref{eq:noise_FPT_def}), (\ref{eq:mean_Nk}), (\ref{eq:CV2_Nk}), and (\ref{eq:MFPT_CB}), and symbols show stochastic simulation results \cite{gillespie1977exact}. Parameter values for these plots: \(\langle B\rangle=1\), \(\langle x_s\rangle=50\) and $\mu = 0.1$, and $k = \langle x_s\rangle\mu N/\langle B\rangle$ to keep $\langle x_s\rangle$ fixed. 
}
\label{fig:FPT_Erlang}
\end{figure}

We first examine how molecular memory in the burst-arrival process affects mRNA fluctuations and, consequently, first-passage-time statistics.
A simple way to introduce such memory is to represent the interburst cycle by $N$ sequential transcriptional states, $D_0\to D_1\to\cdots\to D_{N}\to D_0+{\rm mRNA}$, as shown schematically in Fig.~\ref{fig:FPT_Erlang}(a). 
Here, $D_i$ denotes the $i^{\rm{th}}$ transcriptional state, and the completion of the final transition produces a burst of $B$ mRNAs, which undergo single-step degradation at rate $\mu$.
For $N$ sequential exponential steps, each with the same rate $k$, the interburst time follows an Erlang distribution with mean $N/k$, and coefficient of variation squared $1/N$.
The states $D_i$ are coarse-grained kinetic states, not necessarily distinct molecular species. 
Depending on the biological system, they may capture delays associated with promoter remodeling, assembly of the transcriptional machinery, and the transition to productive transcription~\cite{Weidemann2023SubPoissonian,szavitsnossan2024solving}.
The equal-rate irreversible chain is therefore an effective model of the aggregate waiting time and does not imply that the underlying biochemical steps are identical or map one-to-one onto the $D_i$ states.



To obtain the transient abundance statistics, we derive the moment equations from the joint chemical master equation for the transcriptional state and the mRNA copy number. The detailed calculation is provided in Appendix Sec.~\ref{app:transient_multistep_arrival}. 
The resulting mean and noise (coefficient of variation squared) dynamics are given as follows:
\begin{equation}
\label{eq:mean_Nk}
    \langle x(t) \rangle = \langle x_s\rangle\Big(1-e^{-\mu t}\Big), \qquad \langle x_s\rangle = \frac{k\langle B \rangle}{\mu N}
\end{equation}
\begin{eqnarray}
CV_x^2(t) &=& \frac{1}{2\langle x_s \rangle}\left[1 + \frac{\langle B^2\rangle}{\langle B\rangle}+2\langle B\rangle\left(\frac{1}{(1+\mu/k)^N-1}-\frac{\langle x_s\rangle}{\langle B\rangle}\right)\right]\left(\frac{1+e^{-\mu t}}{1-e^{-\mu t}}\right)-\frac{e^{-\mu t}}{\langle x(t)\rangle}.
\label{eq:CV2_Nk}
\end{eqnarray} 
In the steady-state limit $(t{\to}\infty)$, our noise expression $CV^2_x$ becomes
\begin{equation}
    CV^2_{x_s} = \frac{1}{2\langle x_s \rangle}\left[1 + \frac{\langle B^2\rangle}{\langle B\rangle}+2\langle B\rangle\left(\frac{1}{(1+\mu/k)^N-1}-\frac{\langle x_s\rangle}{\langle B\rangle}\right)\right],
    \label{eq:CV2_Nk_SS}
\end{equation}
which agrees with the previous queueing-theoretic results for bursty gene expression with Erlang-distributed interburst times~\cite{liu1990gix,jia2011intrinsic,kumar2015transcriptional, vahdat2024quantifying}.
For a fixed mean interburst time, hence for a fixed steady-state value of the mean mRNA number ($x_s$), the mean dynamics in Eq.~\eqref{eq:mean_Nk} do not depend on the number of burst arrival steps $N$ (see Fig.~\ref{fig:FPT_Erlang} d, i, respectively, for constant and geometric burst distributions). 
However, the noise in the mRNA level in Eq.~(\ref{eq:CV2_Nk_SS}) decreases with $N$ for any burst distribution. To compare the mean and noise dynamics of mRNA levels with stochastic simulation, we consider constant and a geometric burst size distribution (see  Fig.~\ref{fig:FPT_Erlang} d, e, i, and j).
For constant burst size, each production event generates a fixed number of mRNAs, $B=\langle B\rangle$, such that $\langle B^2\rangle=\langle B\rangle^2$. In particular, $B=1$ corresponds to non-bursty production, where each event produces a single mRNA.
For geometrically distributed burst sizes, we consider the zero-inclusive geometric distribution, $\Pr(B=b)=p(1-p)^b$, for $b=0,1,2,\ldots$, where $p=1/(1+\langle B\rangle)$ and $\langle B^2\rangle=\langle B\rangle(1+2\langle B\rangle)$.

Using the mean mRNA dynamics in Eq.~(\ref{eq:mean_Nk}), the mean FPT, the average time to reach a threshold mRNA level $X$, can be obtained from Eq.~\ref{eq:mean_FPT_def} as
\begin{equation}
    \label{eq:MFPT_CB}
    \langle T \rangle \approx -\frac{1}{\mu}\ln\left(1-\frac{X}{\langle x_s\rangle}\right).
\end{equation}
The mean FPT for a given $X$ is independent of burst arrival steps $N$, which agrees with exact stochastic simulation as well for both constant and geometric burst distributions (see Fig.~\ref{fig:FPT_Erlang} f, k).

Since the FPT noise in Eq.~(\ref{eq:noise_FPT_def}) depends directly on the mRNA noise, the reduction in copy-number fluctuations is expected to improve the precision of threshold-crossing times. This effect is illustrated in Fig.~\ref{fig:FPT_Erlang}(b,c), where sample trajectories cross the same threshold with substantially less temporal dispersion for $N=10$ than for $N=1$. The corresponding timing histogram is narrower for larger $N$, indicating reduced FPT variability. The quantitative dependence of FPT noise on the number of burst-arrival steps is shown in Fig.~\ref{fig:FPT_Erlang} for deterministic and geometric bursts, in panels (g) and (l), respectively. For both cases, increasing $N$ lowers $CV_T^2$.
However, the dependence of the optimal threshold on $N$ is strongly influenced by the burst-size statistics. Figs.~\ref{fig:FPT_Erlang}(h) and (m) show the optimal threshold $X_0$, defined as the value of the normalized threshold $X/\langle x_s\rangle$ at which $CV_T^2$ is minimized, as a function of $N$ for constant and geometric bursts, respectively. For constant burst, increasing the number of transcription steps $N$ shifts $X_0$ below the classical value near $0.5$. In contrast, for geometrically distributed bursts, increasing $N$ lowers the overall magnitude of the FPT noise while leaving the optimal threshold close to one half of the steady-state mean.

So far, we have considered a fixed mean burst size, $\langle B\rangle=1$. To examine how burst magnitude modifies FPT statistics, we vary both $\langle B\rangle$ and $N$ for constant and geometric burst-size distributions in Appendix Fig.~\ref{fig:B_N_plane}. For both distributions, increasing $\langle B\rangle$ increases FPT noise, $\min(CV_T^2)$, which is also consistent with previous FPT studies \cite{Ghusinga2017}.
At a given $N$ and $\langle B\rangle$, geometric bursts produce larger FPT noise than constant burst because of the additional fluctuations in burst size. 
The optimal threshold $X_0$ also depends on mean burst size, $\langle B\rangle$, for both constant and geometric burst distributions. In particular, $X_0$ shifts toward higher values as $\langle B\rangle$ is increased. Thus, the effects of $N$ and $\langle B\rangle$ compete: increasing the number of burst-arrival steps reduces FPT noise and tends to shift the optimal threshold downward, whereas increasing the burst size enhances FPT noise and shifts the optimum toward higher thresholds. These results extend the fixed-$\langle B\rangle$ findings and show that the precision of threshold-crossing times and the location of the optimal threshold are jointly controlled by burst-arrival regularity and burst magnitude. 

\section{Multistep Degradation Reduces Timing Noise}
\label{sec:multi_step_degradation}
\begin{figure}[ht!]
    \centering
    \includegraphics[width=1\linewidth]{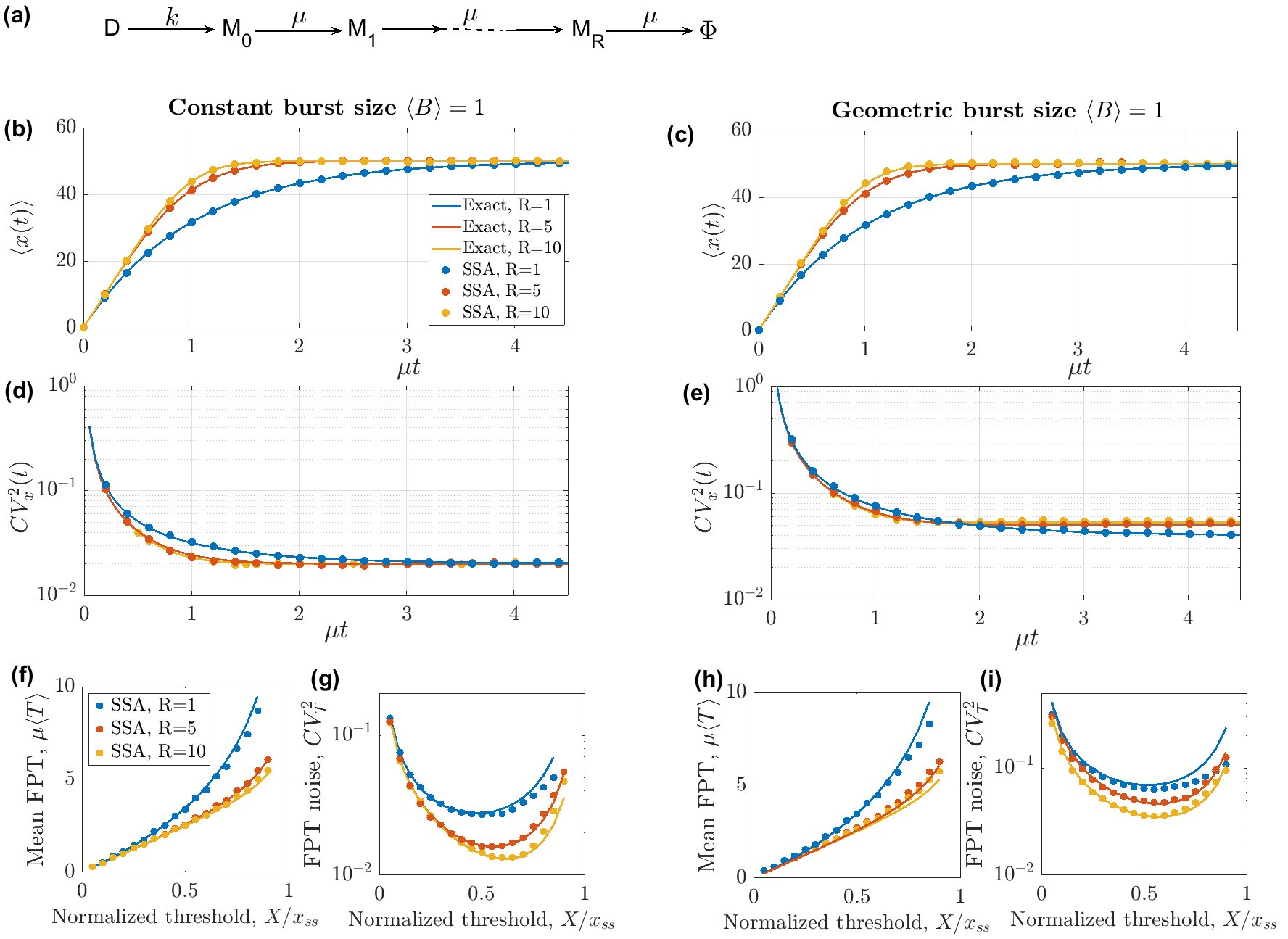}
    \caption{\textbf{For bursty production, multistep mRNA degradation increases copy-number noise but reduces FPT noise.} \textbf{(a)} Schematic of the multistep-degradation model. Bursts arrive with rate $k$ and average size $\langle B\rangle$, and each mRNA passes through $R$ sequential degradation steps, each with transition rate $\mu$, before removal. The total mRNA abundance, $x(t)$, is the sum over all degradation states. \textbf{(b,c)} Mean mRNA dynamics and \textbf{(d,e)} corresponding copy-number noise, $CV_x^2(t)$, for constant and geometric burst size distributions, respectively, with $\langle B\rangle=1$. Solid curves show the exact analytical results in Eqs.~(\ref{eq:mean_total_mrna}) and (\ref{eq:exact_transient_variance}), and symbols denote stochastic simulation algorithm (SSA) \cite{gillespie1977exact} results for $R=1$, $5$, and $10$. \textbf{(f,h)} Mean FPT as a function of the normalized threshold $X/\langle x_s\rangle$. Solid curves show the threshold-crossing estimates obtained from the exact mean dynamics using Eq.~(\ref{eq:mean_FPT_def}), while symbols denote SSA results. \textbf{(g,i)} FPT noise, $CV_T^2$, obtained from SSA matches well with Eq.~(\ref{eq:noise_FPT_def}) under small noise approximation. We keep the mean mRNA lifetime ($R/\mu$) fixed with $R$ by changing $\mu$. The parameter values are $\langle x_s\rangle=50$ and $\mu=R/10$ to keep the mean mRNA lifetime fixed at 10 minutes. The burst-arrival rate is chosen as $k=\langle x_s\rangle\mu/(R\langle B\rangle)$ to keep the steady-state mean fixed.
}
    \label{fig:multi_step_degradation}
\end{figure}

Having considered molecular memory in burst arrival, we next investigate memory generated by multistep mRNA degradation, as illustrated in Fig.~\ref{fig:multi_step_degradation}(a). Standard coarse-grained models represent mRNA degradation as a single first-order reaction, resulting in an exponentially distributed and memoryless transcript lifetime. In cells, however, mRNA degradation can involve several successive biochemical processes, including deadenylation, decapping, and exonucleolytic decay. Such pathways can produce nonexponential, age-dependent lifetime distributions~\cite{Deneke2013ComplexDegradation}. The resulting molecular memory can substantially modify gene-expression fluctuations, as demonstrated in theoretical and experimental studies of general lifetime distributions and multistage gene-expression models~\cite{Pedraza2008MolecularMemory, Park2018ChemicalFluctuation, Weidemann2023SubPoissonian}.

In our model, transcriptional bursts arrive according to a Poisson process with rate $k$ and burst size $B$. Newly produced transcripts enter the first degradation state, $M_0$, and subsequently pass through $R$ sequential states, $M_0,\ldots,M_{R-1}$, before removal. All transitions between degradation states, including removal from the final state, occur with rate $\mu$. The total mRNA abundance $x$ includes transcripts in all degradation states.
For equal transition rates, the mRNA lifetime follows an Erlang distribution with mean mRNA lifetime $R/\mu$, and coefficient of variation squared $1/R$. The case $R=1$ recovers the conventional exponential lifetime, whereas $R>1$ produces a nonexponential lifetime with increasing temporal regularity. When all degradation steps have identical rate $\mu$, the exact mean dynamics of the total mRNA abundance is given as follows:
\begin{align}
\langle x(t)\rangle
&=
\frac{k\langle B\rangle}{\mu}
\left[
R-e^{-\mu t}
\sum_{j=0}^{R-1}
(R-j)\frac{(\mu t)^j}{j!}
\right].
\label{eq:mean_total_mrna}
\end{align}
The first term represents the steady-state mean mRNA abundance, determined by the mean production rate and the average mRNA lifetime. The second term accounts for the transient buildup from the initially empty system, with its time dependence reflecting the sequential degradation steps. As this transient contribution vanishes at long times, the mean mRNA abundance approaches its steady-state value,
\begin{equation}
\langle x_s\rangle=\frac{k\langle B\rangle R}{\mu}.
\label{eq:mean_total_mrna_ss}
\end{equation}
The detailed derivation is provided in Appendix Sec.~\ref{app:mean_multistep_decay}.
The mean mRNA dynamics for constant and geometric burst distributions are shown, respectively, in Fig.~\ref{fig:multi_step_degradation}(b) and (c) for $\langle B\rangle = 1$. For both burst distributions, Eq.~(\ref{eq:mean_total_mrna}) agrees well with the stochastic simulations for $R=1$, $5$, and $10$ degradation steps. When the mean mRNA lifetime $R/\mu$ is held fixed, increasing $R$ makes the dynamics faster. Consequently, the mean FPT defined in Eq.~(\ref{eq:mean_FPT_def}) reduces with $R$, as shown in Fig.~\ref{fig:multi_step_degradation}(f,h). The analytical predictions for the mean FPT, obtained using Eq.~(\ref{eq:mean_FPT_def}), also agree well with the stochastic simulations for multistep mRNA degradation.

The noise in the mRNA level is quantified by the squared CV, which is the ratio between the variance and the square of the mean. For the multistep-degradation model, we derive the noise dynamics in Appendix Sec~\ref{app:variance_multistep_decay}, given as follows:
\begin{equation}
CV^2_x(t) = \frac{1}{\langle x(t)\rangle} + \left(\langle B^2\rangle-\langle B\rangle\right) \frac{k}{\mu \langle x(t)\rangle^2}
\sum_{i=0}^{R-1}
\sum_{j=0}^{R-1}
\frac{(i+j)!}{i!j!2^{i+j+1}}
\times
\left[
1-e^{-2\mu t}
\sum_{\ell=0}^{i+j}
\frac{(2\mu t)^\ell}{\ell!}
\right].
\label{eq:exact_transient_variance}
\end{equation}
The first term represents the Poisson contribution to mRNA fluctuations, whereas the second captures the additional fluctuations associated with bursty production. The magnitude of the latter depends on both the burst-size statistics and the degradation dynamics, which determine how long transcripts produced within the same burst survive together.
Taking the long-time limit ($t \to \infty$), the steady-state noise becomes
\begin{equation}
CV^2_{x_s}
=
\frac{1}{\langle x_s\rangle}
\left(
1+
\frac{
\langle B^2\rangle-\langle B\rangle
}{
\langle B\rangle
}
q_R
\right), \qquad q_R=1-\frac{\binom{2R}{R}}{4^R},
\label{eq:steady_state_variance_general}
\end{equation}
which agrees with the general queueing-theoretic expression~\cite{jia2011intrinsic} when the molecular lifetime distribution is specialized to the Erlang distribution arising from $R$ sequential degradation steps.
The factor $q_R$ captures the effect of multistep degradation, which increases monotonically with $R$, hence increasing the steady-state mRNA noise. For unit-sized bursts ($B=1$), this additional contribution vanishes, and the steady-state abundance remains Poissonian, independently of the number of degradation steps $R$~\cite{szavitsnossan2024solving}. 
We validate the noise dynamics in mRNA levels for multi-step mRNA degradation with stochastic simulation in Fig.~\ref{fig:multi_step_degradation} (d) and (e), respectively, for constant and geometric burst size distributions. 
For constant bursts ($\langle B^2\rangle = \langle B\rangle^2$), with unit burst size, \(\langle B\rangle=1\), Eq.~\ref{eq:steady_state_variance_general} shows that the steady-state noise is Poissonian, where $CV_{x_s}^2=1/\langle x_s\rangle$, which is independent of \(R\). 
However, for \(\langle B\rangle>1\), $CV_{x_s}^2$ increases with \(R\) at fixed mean mRNA lifetime.
However, for geometrically distributed bursts, \(\langle B^2\rangle=\langle B\rangle(1+2\langle B\rangle)\); consequently, Eq.~\ref{eq:steady_state_variance_general} predicts that the steady-state noise increases monotonically with \(R\) for any \(\langle B\rangle>0\) at fixed \(\langle x_s\rangle\).
For the same mean burst size $\langle B\rangle$, geometric bursts produce larger fluctuations than constant ones because of their additional burst-size variability, which is also evident in Fig.~\ref{fig:multi_step_degradation} (d) and (e).

In contrast to the behavior of the steady-state abundance noise, the first-passage-time noise decreases as the number of degradation steps, $R$, increases, as shown by the stochastic simulation results in Fig.~\ref{fig:multi_step_degradation}(g) and (i), respectively, for constant and geometric burst distributions. This reduction arises through a mechanism distinct from multistep burst arrival: at fixed mean mRNA lifetime, increasing the number of degradation steps accelerates the mean mRNA response and increases the logarithmic sensitivity $S_x$ in Eq.~(\ref{eq:noise_FPT_def}), thereby contributing to lower FPT noise. Thus, degradation memory can improve timing precision even when it does not reduce steady-state abundance noise.

The behavior of the optimal threshold further distinguishes multi-step degradation from the multi-step burst arrival process. For multi-step degradation, $X_0$ remains close to or above the classical value of $0.5$ over most of the parameter range examined and shifts toward higher values as $\langle B\rangle$ increases (see Appendix Fig.~\ref{fig:R_B_plane} g, j). At the smallest burst sizes and largest values of $R$, the optimal threshold remains close to $0.5$. Thus, unlike multistep burst arrival, multistep degradation does not produce a systematic shift of $X_0$ toward lower thresholds. Degradation memory therefore primarily improves temporal precision while preserving a relatively high optimal threshold, whereas burst-arrival memory can modify both FPT precision and the location of the optimal threshold.

\section{Combined Effects of Multistep Transcription and Degradation}
\label{sec:combined_memory}

\begin{figure}[ht!]
\centering
\includegraphics[width=0.9\linewidth]{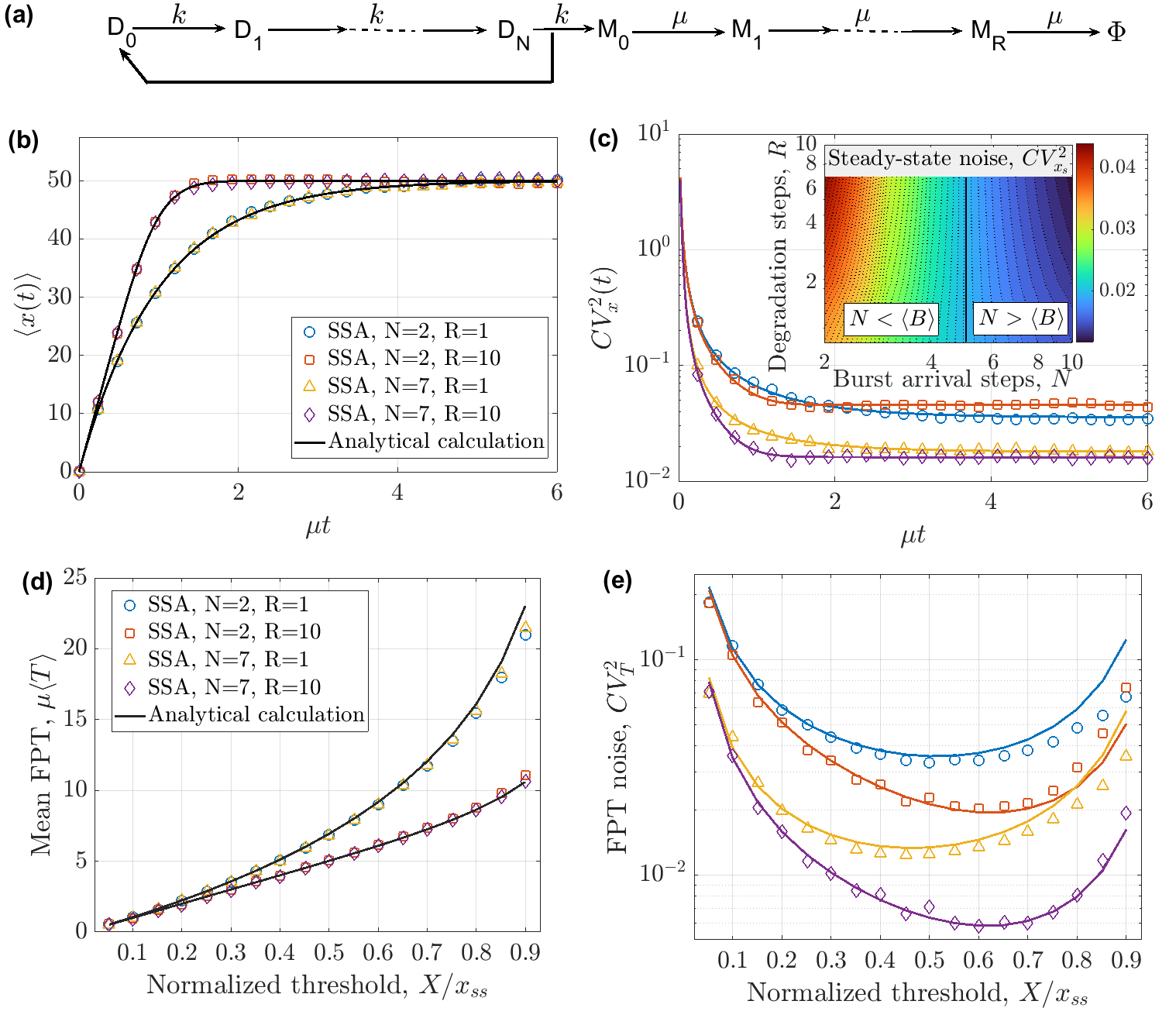}
\caption{\textbf{Combined effects of multistep transcription and mRNA degradation on FPT noise for constant burst.}
Symbols denote SSA results; curves show analytical predictions or
approximations as specified below.
\textbf{(a)} Schematic of reaction scheme. Transcription proceeds through $N$ sequential stages at rate $k$, producing a deterministic mRNA burst upon completion and resetting the cycle. Each transcript independently traverses $R$ degradation stages at rate $\mu$ before removal.
\textbf{(b)} Mean mRNA dynamics for $(N,R)=(2,1),(2,10),(7,1)$, and $(7,10)$. Black curves show Eq.~(\ref{eq:mean_combined_dynamics}). The mean dynamics is independent of $N$ and approaches the common steady state more rapidly as $R$ increases.
\textbf{(c)} Transient mRNA noise, $CV_x^2(t)$, which decreases with increasing $N$. Solid curves show Eq.~(\ref{eq:CV2_combined_dynamics}).
Inset: steady-state noise in the $N-R$ plane for $\langle B\rangle = 5$. Increasing $N$ reduces noise for all ranges of $R$. However, increasing $R$ raises the steady-state noise for $N<\langle B\rangle$ and lowers it for $N>\langle B\rangle$.
\textbf{(d)} Mean FPT as a function of the normalized threshold $X/\langle x_s\rangle$. Solid curves show the threshold-crossing estimates obtained from the exact mean dynamics using Eq.~(\ref{eq:mean_FPT_def}). The predicted mean FPT is independent of $N$ at fixed $R$, so the corresponding analytical curves overlap. Increasing $R$ reduces the mean FPT independent of threshold values.
\textbf{(e)} FPT noise, $CV_T^2$, versus normalized threshold. Curves show the small-noise approximation, Eq.~(\ref{eq:noise_FPT_def}), evaluated using the exact transient mRNA mean and noise. FPT noise is nonmonotonic in the threshold, with a minimum whose value and location depend on $N$ and $R$. Increasing both $N$ and $R$ reduces FPT noise. 
All results use $\langle B\rangle=5$, $\mu=R/10$, and $k=N$, fixing the mean interburst time at $N/k=1$ and the mean mRNA lifetime at $R/\mu=10$ and mean steady-state mRNA number $\langle x_s\rangle = (\langle B\rangle k R)/(\mu N) = 50$.
}
\label{fig:combined_memory}
\end{figure}

We next consider multistep burst arrival and mRNA degradation within a single model [Fig.~\ref{fig:combined_memory}(a)]. Burst production requires completion of $N$ sequential transcriptional stages, each with rate $k$, whereas each transcript independently traverses $R$ degradation stages, each with rate $\mu$, before removal. Burst sizes are deterministic, $B=\langle B\rangle$, and the mRNA abundance $x$ counts transcripts in all degradation stages. To isolate the effects of changing the number of stages, we hold the mean interburst time, $N/k$, and the mean mRNA lifetime, $R/\mu$, fixed. 

Under these conditions, the exact mean dynamics are
\begin{equation}
    \langle x(t)\rangle
    =\frac{k\langle B\rangle}{N\mu}
    \sum_{j=1}^{R}\left[
        1-e^{-\mu t}\sum_{\ell=0}^{j-1}
        \frac{(\mu t)^\ell}{\ell!}
    \right].
    \label{eq:mean_combined_dynamics}
\end{equation}
The derivation is given in Appendix~\ref{app:steady_state_noise_NR}.
At fixed mean interburst time, increasing $N$ changes the temporal
regularity of burst production without changing its mean rate; hence,
the mean mRNA trajectory is independent of $N$. Increasing $R$ at fixed
mean lifetime narrows the lifetime distribution and accelerates the
approach to the common steady state. These properties are confirmed by
the analytical curves and stochastic simulations in
Fig.~\ref{fig:combined_memory}(b).

Unlike the mean, the transient noise depends on both $N$ and $R$.
Writing the mean burst arrival rate $\lambda=k/N$, the exact squared coefficient of variation for constant burst can be expressed as
\begin{equation}
    CV_x^2(t)=
    \frac{
        \langle x(t)\rangle
        +\lambda\langle B\rangle(\langle B\rangle-1)I_R(t)
        +2\lambda\langle B\rangle^2J_{N,R}(t)
    }{\langle x(t)\rangle^2},
    \qquad t>0.
    \label{eq:CV2_combined_dynamics}
\end{equation}
Here, $I_R(t)=\int_0^t S_R(u)^2\,du$, where $S_R(u)$ is the survival
probability of an individual transcript, and $J_{N,R}(t)$ is the
survival-weighted integral of correlations between distinct burst
arrivals. Their explicit expressions and derivation are given in
Appendix~\ref{app:steady_state_noise_NR}. The three terms in the numerator
represent the individual-molecule contribution, the contribution from
molecules produced in the same burst, and the contribution from
correlations between different bursts, respectively. For Poisson burst
arrivals ($N=1$), $J_{1,R}(t)=0$. Multistep arrival modifies the last term
and suppresses the transient fluctuations in the cases shown in
Fig.~\ref{fig:combined_memory}(c), while leaving the mean trajectory
unchanged. The exact transient prediction agrees with the simulations.

The steady-state mean and the large-copy-number approximation to the
steady-state noise are
\begin{eqnarray}
    \langle x_s\rangle
    &=&\frac{k\langle B\rangle R}{N\mu},
    \label{eq:mean_ss_combined}\\
    CV_{x_s}^2
    &\simeq&\frac{1}{\langle x_s\rangle}
    \left[1+\left(\frac{\langle B\rangle}{N}-1\right)q_R\right],
    \qquad
    q_R=1-\frac{\binom{2R}{R}}{4^R}.
    \label{eq:FF_combined_approx}
\end{eqnarray}
Since $q_R$ increases monotonically with $R$, the effect of multistep degradation depends on the ratio $\langle B\rangle/N$. At fixed $\langle x_s\rangle$, increasing $R$ reduces the steady-state noise when $\langle B\rangle<N$, leaves it unchanged within this approximation when $\langle B\rangle=N$, and increases it when $\langle B\rangle>N$. 
The inset of Fig.~\ref{fig:combined_memory}(c) illustrates the opposite trends.
For $N=1$,Eq.~(\ref{eq:FF_combined_approx}) recovers Eq.~(\ref{eq:steady_state_variance_general}) for constant burst, $\langle B^2\rangle=\langle B\rangle^2$.

We then examine the first-passage time to a threshold $X$ in total mRNA abundance. The mean-crossing estimate in Eq.~(\ref{eq:mean_FPT_def}) depends only on the mean trajectory and is therefore independent of $N$ at fixed $R$. Increasing $R$ lowers this estimate, consistent with the faster accumulation of mRNA [Fig.~\ref{fig:combined_memory}(d)]. The simulations follow the predicted trends, with
larger deviations at high thresholds.

The small-noise approximation in Eq.~(\ref{eq:noise_FPT_def}) depends on both the transient mRNA noise and the slope of the mean trajectory at the mean-crossing time. Consequently, steady-state abundance noise alone does not determine FPT noise. As shown in Fig.~\ref{fig:combined_memory}(e), $CV_T^2$ is nonmonotonic in the threshold, and both its minimum and the threshold at which it occurs depend on $N$ and $R$. Increasing $N$ and $R$ reduces FPT noise. Among the combinations considered, $(N,R)=(7,10)$ yields the lowest FPT noise. Deviations from the small-noise prediction become more pronounced at high copy-number noise. These results show that temporal precision reflects both production regularity and the transient response shaped by degradation.

\section{Discussion and Conclusions}
\label{sec:discussion}
We have investigated how molecular memory generated by multistep burst arrival and multistep mRNA degradation affects the precision of threshold-crossing times in stochastic gene expression. Previous studies of Markovian models found that FPT noise is minimized at a threshold close to one-half of the steady-state expression level~\cite{Ghusinga2017FirstPassage, DalCo2017StochasticTiming, Kannoly2020OptimumThreshold}. Our analytical calculations and stochastic simulations show that the optimal threshold is not universal for nonexponential waiting times. Both forms of memory reduce FPT noise, but through distinct mechanisms. At fixed mean burst arrival time, increasing the number of burst-arrival steps regularizes production and suppresses copy-number fluctuations without altering the mean accumulation trajectory. Within the small-noise approximation, this reduction in abundance noise translates into lower FPT noise. Multistep degradation instead alters the transient trajectory and its time sensitivity. When the mean mRNA lifetime is fixed, increasing the number of degradation steps accelerates the mean mRNA response and increases the logarithmic sensitivity, thereby contributing to lower FPT noise. For constant burst, production memory also shifts the optimal normalized threshold below one-half, whereas degradation memory generally leaves the optimum close to or above one-half. Thus, production and degradation memory can reinforce one another in reducing FPT noise while having distinct, and in some regimes opposing, effects on the location of its minimum.

The predicted dependence of timing precision on molecular memory is experimentally testable. Live-cell studies have revealed substantial variation in transcriptional bursting kinetics among mammalian genes, and recent work has emphasized the importance of resolving the relevant timescales underlying transcriptional bursting \cite{Suter2011BurstingKinetics,Meeussen2024TimeWillTell}. For an inducible reporter whose expression activates a tunable downstream threshold, one could simultaneously estimate burst waiting-time distributions, burst-size distributions, and threshold-crossing-time statistics. The present theory predicts that promoters with more regular burst arrival should exhibit reduced FPT noise; moreover, a shift of the optimal threshold below one-half should be most apparent when burst-size fluctuations are small. Phage $\lambda$ provides a complementary test bed in which the holin threshold and translational efficiency can be perturbed while lysis times are measured in single cells~\cite{Kannoly2020OptimumThreshold,Kannoly2025TranslationEfficiency}. Synthetic manipulation of RNA processing or degradation pathways could further test the prediction that degradation memory improves timing precision without reproducing the same threshold shift as production memory.

The analytical treatment relies on a small-noise approximation and on a regime in which burst arrival is fast compared with mRNA degradation. These assumptions allow the FPT noise to be related directly to transient abundance fluctuations and the slope of the mean trajectory. Deviations between the analytical approximation and stochastic simulations are expected when burst sizes become large, when burst-amplitude fluctuations are strong, or when threshold crossing is dominated by discrete overshoot events. Nevertheless, the simulations show that the qualitative conclusions persist beyond the strict approximation regime: multistep arrival regularizes timing, burst-size statistics determine whether the optimal threshold shifts, and multistep degradation provides an additional route to lower timing variability. These passive kinetic-memory effects are complementary to active timing control by feedforward or feedback circuits and to robustness against external disturbances~\cite{Dey2021Feedforward,Vahdat2021Feedback,Rezaee2023SelfRegulated,Rezaee2024RegulatedDegradation,GhusingaSingh2020ExternalDisturbances}. Extensions incorporating explicit promoter switching, protein-level thresholds, extrinsic noise, cell division and dilution, or correlated production and degradation memory would further clarify how non-Markovian gene-expression dynamics regulate cellular decision times.

In summary, molecular memory is not merely a correction to Markovian models of gene expression; it provides a mechanism for controlling temporal precision. Production memory suppresses fluctuations in stochastic input and can shift the optimal threshold to earlier expression levels when bursts are sufficiently regular, whereas degradation memory improves timing through modified transcript persistence and transient accumulation. These findings establish that the precision and optimal design of intracellular timing are determined not only by mean expression levels, but also by the temporal organization and amplitude statistics of molecular events.
%


\appendix

\section{Transient Mean and Variance for Multistep Burst Arrival}
\label{app:transient_multistep_arrival}

\begin{figure}[ht!]
\centering
\includegraphics[width=1\linewidth]{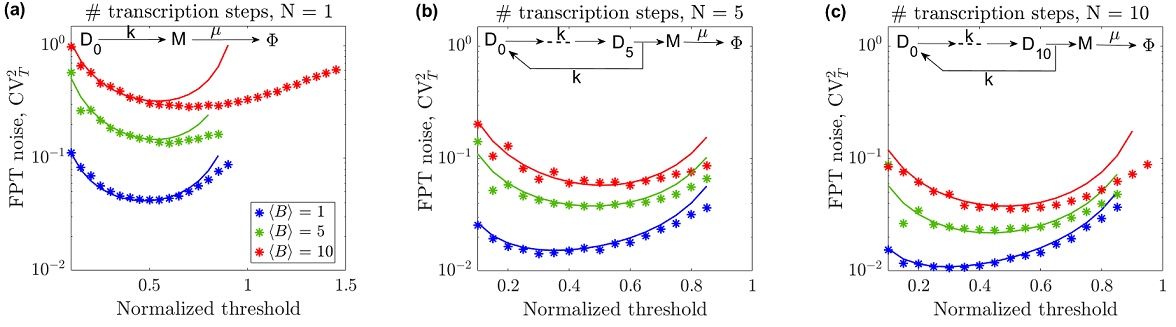}\
\vspace{0.5 cm}
\includegraphics[width=1\linewidth]{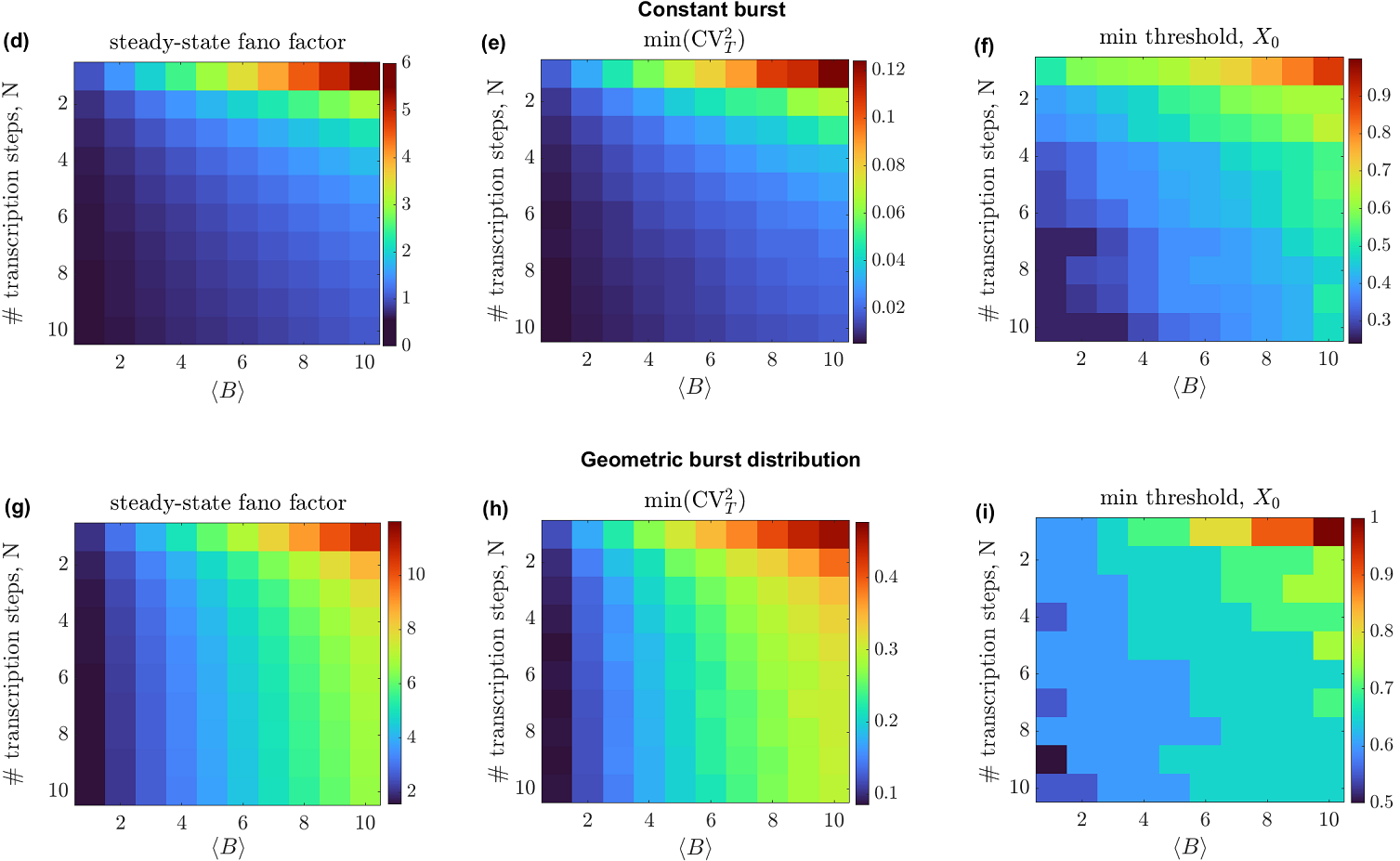}
\caption{\textbf{FPT noise with $\langle B\rangle$  and $N$ for constant burst and multi-step transcription.} FPT noise with normalized threshold ($X/\langle x_s\rangle$) for different mean burst size, $\langle B\rangle$ when burst is constant and number of transcription steps, \textbf{(a)} $N$ = 1, \textbf{(b)} $N$ = 5, and \textbf{(c)} $N$ = 10. In the large noise limit, our approximated SNA calculation (solid lines Eq.~\ref{eq:noise_FPT_def} does not hold with SSA simulation (symbols) \cite{gillespie1977exact}. \textbf{(d)} Steady-state Fano factor of mRNA level, \textbf{(e)} minima value of FPT noise, and \textbf{(f)} minima value of normalized threshold in the $\langle B\rangle-N$ plane, obtained from SSA simulation \cite{gillespie1977exact} over $10^4$ realizations. \textbf{(g)-(i) Geometric burst distribution:} the same analysis is presented in the last row for the geometric burst distribution, where $X_0 \gtrsim 0.5$, unlike the constant burst case, where $X_0$ can fall below 0.5.
Parameter values: $\langle x_s\rangle$ = 50, $\langle\tau\rangle$ = 2, $k = N/\langle\tau\rangle$, and $\mu = \langle B\rangle/(\langle x_s\rangle \langle\tau\rangle)$ are tuned to keep $\langle x_s\rangle$ and $\langle\tau\rangle$ fixed for different $N$ and $\langle B\rangle$.
}
\label{fig:B_N_plane}
\end{figure}

We derive the transient mean and variance of the mRNA abundance for the multistep burst-arrival model considered in Sec.~\ref{sec:multi_step_arrival}. The transcriptional cycle consists of $N$ sequential states,
\begin{equation}
D_0 \xrightarrow{k} D_1 \xrightarrow{k}
\cdots \xrightarrow{k} D_{N-1}
\xrightarrow{k} D_0+B M,
\label{app:eq:multistep_arrival_scheme}
\end{equation}
where completion of the final transition produces a burst of mRNAs.
Individual mRNAs undergo first-order degradation,
\begin{equation}
M\xrightarrow{\mu}\varnothing.
\end{equation}
We denote the
burst-size distribution by
\begin{equation}
\rho_b=\Pr(B=b),
\qquad b=0,1,2,\ldots,
\label{app:eq:burst_distribution}
\end{equation}
with moments
\begin{equation}
\langle B\rangle
=
\sum_{b=0}^{\infty}b\rho_b,
\qquad
\langle B^2\rangle
=
\sum_{b=0}^{\infty}b^2\rho_b.
\end{equation}
The chemical master equation for state $D_0$ is
\begin{align}
\frac{\partial P_0(x,t)}{\partial t}
={}&
k\sum_{b=0}^{\infty}
\rho_b P_{N-1}(x-b,t)
-kP_0(x,t) +
\mu(x+1)P_0(x+1,t)
-\mu xP_0(x,t),
\label{app:eq:master_equation_D0}
\end{align}
and, for the remaining transcriptional states ($i=1,2,..,N-1$),
\begin{align}
\frac{\partial P_i(x,t)}{\partial t}
={}&
kP_{i-1}(x,t)-kP_i(x,t)
+
\mu(x+1)P_i(x+1,t)
-\mu xP_i(x,t),
\label{app:eq:master_equation_Di}
\end{align}
where
\begin{equation}
P_i(x,t)
=
\Pr\!\left[x(t)=x,\;D_i(t)=1\right],
\qquad
i=0,\ldots,N-1,
\label{app:eq:joint_probability}
\end{equation}
denote the joint probability that there are $x$ mRNAs and that the transcriptional cycle is in state $D_i$ at time $t$. The probability of occupying transcriptional state $D_i$ is
\begin{equation}
p_i(t)
=
\langle D_i(t)\rangle
=
\sum_{x=0}^{\infty}P_i(x,t),
\label{app:eq:state_probability_from_master}
\end{equation}
and the moments of the total mRNA abundance are
\begin{equation}
\langle x^n(t)\rangle
=
\sum_{i=0}^{N-1}
\sum_{x=0}^{\infty}
x^nP_i(x,t).
\label{app:eq:moments_from_master}
\end{equation}
The state probabilities satisfy
\begin{align}
\frac{dp_0}{dt}
&=
kp_{N-1}-kp_0,
\\
\frac{dp_i}{dt}
&=
kp_{i-1}-kp_i,
\qquad i=1,\ldots,N-1.
\label{app:eq:state_probability}
\end{align}
Let $D_i(t)\in\{0,1\}$ denote the occupancy of transcriptional state $i$,
such that
\begin{equation}
\sum_{i=0}^{N-1}D_i(t)=1,
\end{equation}
and we obtain 
\begin{equation}
p_i(t)=\frac{1}{N},
\label{app:eq:uniform_state_probability}
\end{equation}
by assuming burst arrivals are faster than the mRNA lifetime.
\subsection{Exact mean dynamics of mRNA}

Let $m(t)=\langle x(t)\rangle$ denote the mean total mRNA abundance. A burst-production event changes
$x$ by $B$, whereas a degradation event changes $x$ by $-1$. The exact
mean equation is therefore
\begin{equation}
\frac{dm}{dt}
=
k\langle B\rangle p_{N-1}(t)-\mu m(t).
\label{app:eq:exact_mean_general_phase}
\end{equation}
For an arbitrary initial transcriptional phase and $m(0)=0$,
Eq.~\eqref{app:eq:exact_mean_general_phase} gives
\begin{equation}
m(t)
=
k\langle B\rangle
\int_0^t
p_{N-1}(t')e^{-\mu(t-t')}\,dt'.
\label{app:eq:mean_arbitrary_phase}
\end{equation}
Thus, the transient mean generally depends on the initialization of the
transcriptional cycle. For the stationary phase initialization in
Eq.~\eqref{app:eq:uniform_state_probability}, the mean equation
reduces to
\begin{equation}
\frac{dm}{dt}
=
\frac{k\langle B\rangle}{N}-\mu m(t).
\label{app:eq:mean_stationary_phase}
\end{equation}
Starting from an empty mRNA population, $m(0)=0$, its exact solution is
\begin{equation}
\boxed{
\langle x(t)\rangle
=
\frac{k\langle B\rangle}{\mu N}
\left(1-e^{-\mu t}\right)
}
\label{app:eq:exact_mean_multistep_arrival}
\end{equation}
where the steady-state mean mRNA number in the limit $t \to \infty$ is
\begin{equation}
\boxed{
\langle x_s\rangle
=
\frac{k\langle B\rangle}{\mu N}
}.
\label{app:eq:mean_ss_multistep_arrival}
\end{equation}

\subsection{Exact second-moment and variance equations}


The exact differential equation for the second moment is
\begin{align}
\frac{d\langle x^2\rangle}{dt}
={}& 2k\langle B\rangle \langle xD_{N-1}\rangle+k\langle B^2\rangle \langle D_{N-1}\rangle-2\mu\langle x^2\rangle+\mu\langle x\rangle.
\label{app:eq:exact_second_moment}
\end{align}
Defining
\begin{equation}
V(t)=\operatorname{Var}[x(t)]
=
\langle x^2(t)\rangle-\langle x(t)\rangle^2,
\end{equation}
and using Eq.~\eqref{app:eq:exact_mean_general_phase}, we obtain
\begin{equation}
\frac{dV}{dt}
=
-2\mu V
+\mu \langle x\rangle
+k\langle B^2\rangle \langle D_{N-1}\rangle
+
2k\langle B\rangle
\operatorname{Cov}(x,D_{N-1}).
\label{app:eq:exact_variance_covariance}
\end{equation}
\subsection{Steady-state covariance and abundance noise}
To obtain variance dynamics from Eq.~(\ref{app:eq:exact_variance_covariance} we need to know the mixed moment between the mRNA abundance and the burst-producing transcriptional state. We define
\begin{equation}
u_i(t)=\langle x(t)D_i(t)\rangle,
\qquad i=0,\ldots,N-1.
\label{app:eq:mixed_moment_definition}
\end{equation}
These mixed moments obey the closed linear system
\begin{align}
\frac{du_0}{dt}
&=
-(k+\mu)u_0
+ku_{N-1}
+k\langle B\rangle \langle D_{N-1}\rangle,
\label{app:eq:mixed_moment_0}
\\
\frac{du_i}{dt}
&=
ku_{i-1}-(k+\mu)u_i,
\qquad i=1,\ldots,N-1.
\label{app:eq:mixed_moment_i}
\end{align}
For an initially empty mRNA population,
\begin{equation}
u_i(0)=0,
\qquad i=0,\ldots,N-1.
\end{equation}
For the stationary phase distribution, $\langle D_i\rangle=1/N$.
Setting the time
derivatives in Eqs.~\eqref{app:eq:mixed_moment_0} and
\eqref{app:eq:mixed_moment_i} to zero gives the steady-state solution
\begin{equation}
u_{i,s}
=
\left(\frac{k}{k+\mu}\right)^i u_{0,s}.
\end{equation}
The mixed moment involving the burst-producing state is
\begin{equation}
u_{N-1,s}
=
\frac{\langle B\rangle}{N}
\frac{1}
{\left(1+\mu/k\right)^N-1}.
\label{app:eq:mixed_moment_ss}
\end{equation}
Since
\begin{equation}
\langle x_s\rangle \langle D_{N-1}\rangle
=
\frac{k\langle B\rangle}{\mu N^2},
\end{equation}
the steady-state covariance is
\begin{equation}
\boxed{
\operatorname{Cov}(x,D_{N-1})_s
=
\frac{\langle B\rangle}{N}
\left[
\frac{1}{(1+\mu/k)^N-1}
-\frac{k}{\mu N}
\right].
}
\label{app:eq:covariance_ss}
\end{equation}
For $N=1$, this covariance vanishes. For $N>1$, it is negative because
\begin{equation}
(1+\mu/k)^N-1>N\mu/k.
\end{equation}
This negative correlation between the mRNA abundance and the
burst-producing state contributes to the suppression of mRNA
fluctuations by multistep burst arrival.
Using Eq.~\eqref{app:eq:covariance_ss} in the steady-state limit of
Eq.~\eqref{app:eq:exact_variance_covariance} gives
\begin{equation}
CV_{x_s}^2
=
\frac{1}{2\langle x_s\rangle}
\left[
1+
\frac{\langle B^2\rangle}{\langle B\rangle}
+
2\langle B\rangle\left(\frac{1}{(1+\mu/k)^N-1}-\frac{k}{\mu N}\right)
\right],
\label{app:eq:CV_ss_multistep_arrival}
\end{equation}


\subsection{Fast-arrival approximation for the transient variance}

A closed approximation for the variance dynamics can be obtained when all nonzero relaxation modes of the burst-arrival process are fast compared with mRNA degradation. 
After the short arrival-process transient, we approximate the combined production and covariance contribution in
Eq.~\eqref{app:eq:exact_variance_covariance} by its steady-state value:
The approximate transient variance equation is therefore
\begin{equation}
\frac{dV}{dt}
\approx
-2\mu V
+\mu\langle x(t)\rangle
+2\mu V_s
-\mu\langle x_s\rangle,
\label{app:eq:approximate_variance_ode}
\end{equation}
where
\begin{equation}
V_s
=
\operatorname{Var}[x_s]
=
CV_{x_s}^2\langle x_s\rangle^2.
\end{equation}
Using the mean dynamics in
Eq.~\eqref{app:eq:exact_mean_multistep_arrival} and the initial
condition $V(0)=0$, the integrating-factor solution of
Eq.~\eqref{app:eq:approximate_variance_ode} is
\begin{align}
V(t)
&=
V_s\left(1-e^{-2\mu t}\right)
-
\langle x_s\rangle e^{-\mu t}
\left(1-e^{-\mu t}\right).
\label{app:eq:variance_dynamics_multistep_arrival}
\end{align}
Finally, dividing
Eq.~\eqref{app:eq:variance_dynamics_multistep_arrival} by
\begin{equation}
\langle x(t)\rangle^2
=
\langle x_s\rangle^2
\left(1-e^{-\mu t}\right)^2
\end{equation}
gives
\begin{equation}
\boxed{
CV_x^2(t)
\approx
CV_{x_s}^2
\left(
\frac{1+e^{-\mu t}}
{1-e^{-\mu t}}
\right)
-
\frac{e^{-\mu t}}
{\langle x(t)\rangle}
}.
\label{app:eq:CV_dynamics_multistep_arrival}
\end{equation}



\section{Multistep mRNA Degradation}

We derive the steady-state mean mRNA abundance for the multistep-degradation model considered in Sec.~\ref{sec:multi_step_degradation}. Bursts arrive according to a Poisson process with rate $k$. Each burst produces a random number $B$ of mRNAs, with mean $\langle B\rangle$, and all newly produced mRNAs enter the first degradation state, $M_0$. Each transcript then passes through $R$ sequential degradation states,
\begin{equation}
M_0 \xrightarrow{\mu} M_1 \xrightarrow{\mu}
\cdots \xrightarrow{\mu} M_{R-1}
\xrightarrow{\mu} \varnothing .
\end{equation}
The total mRNA abundance is defined as
\begin{equation}
x(t)=\sum_{r=0}^{R-1}M_r(t).
\label{app:eq:total_mrna}
\end{equation}

\subsection{Exact Mean Dynamics}
\label{app:mean_multistep_decay}

Let denote the mean mRNA as
\begin{equation}
m_r(t)=\langle M_r(t)\rangle.
\end{equation}
Because the reaction network is linear, the equations governing the mean
copy numbers close exactly:
\begin{align}
\frac{d m_0}{dt}
&=
k\langle B\rangle -\mu m_0,
\label{app:eq:mean_m0}
\\
\frac{d m_r}{dt}
&=
\mu m_{r-1}-\mu m_r,
\qquad r=1,\ldots,R-1.
\label{app:eq:mean_mr}
\end{align}
We consider the following initial condition
\begin{equation}
m_r(0)=0,
\qquad r=0,\ldots,R-1.
\label{app:eq:empty_initial_condition}
\end{equation}
Denoting the Laplace transform of $m_r(t)$ by
$\widetilde{m}_r(s)$, Eq.~\eqref{app:eq:mean_m0} gives
\begin{equation}
\widetilde{m}_0(s)
=
\frac{k\langle B\rangle }{s(s+\mu)}.
\end{equation}
For the subsequent degradation states,
Eq.~\eqref{app:eq:mean_mr} gives the recursion
\begin{equation}
\widetilde{m}_r(s)
=
\frac{\mu}{s+\mu}\widetilde{m}_{r-1}(s)
=
\frac{k\langle B\rangle \mu^r}{s(s+\mu)^{r+1}}.
\label{app:eq:laplace_mean_mr}
\end{equation}
Inverting Eq.~\eqref{app:eq:laplace_mean_mr}, we obtain
\begin{equation}
m_r(t)
=
\frac{k\langle B\rangle}{\mu}
\left[
1-e^{-\mu t}
\sum_{j=0}^{r}\frac{(\mu t)^j}{j!}
\right],
\qquad r=0,\ldots,R-1.
\label{app:eq:mean_individual_state}
\end{equation}
Summing Eq.~\eqref{app:eq:mean_individual_state} over all degradation states gives the exact mean dynamics of the total mRNA abundance:
\begin{align}
\langle x(t)\rangle
&=
\frac{k\langle B\rangle}{\mu}
\left[
R-e^{-\mu t}
\sum_{j=0}^{R-1}
(R-j)\frac{(\mu t)^j}{j!}
\right].
\label{app:eq:mean_total_mrna}
\end{align}
Taking the long-time limit yields
\begin{equation}
\langle x_s\rangle
=
\lim_{t\rightarrow\infty}\langle x(t)\rangle
=
\frac{k\langle B\rangle R}{\mu}.
\label{app:eq:mean_total_mrna_ss}
\end{equation}
The normalized transient mean can therefore be written as
\begin{equation}
\frac{\langle x(t)\rangle}{\langle x_s\rangle}
=
1-\frac{e^{-\mu t}}{R}
\sum_{j=0}^{R-1}
(R-j)\frac{(\mu t)^j}{j!}.
\label{app:eq:normalized_mean_total_mrna}
\end{equation}
In our simulations, the steady-state mean is held fixed by choosing
\begin{equation}
k=\frac{\mu\langle x_s\rangle}{R\langle B\rangle}.
\label{app:eq:mu_fixed_mean}
\end{equation}
Equations~\eqref{app:eq:mean_total_mrna} and
\eqref{app:eq:normalized_mean_total_mrna} are exact and do not require a small-noise or timescale-separation approximation. Moreover, the burst-size distribution affects the mean dynamics only through $\langle B\rangle$. 
\begin{figure}[ht!]
\centering
\includegraphics[width=1\linewidth]{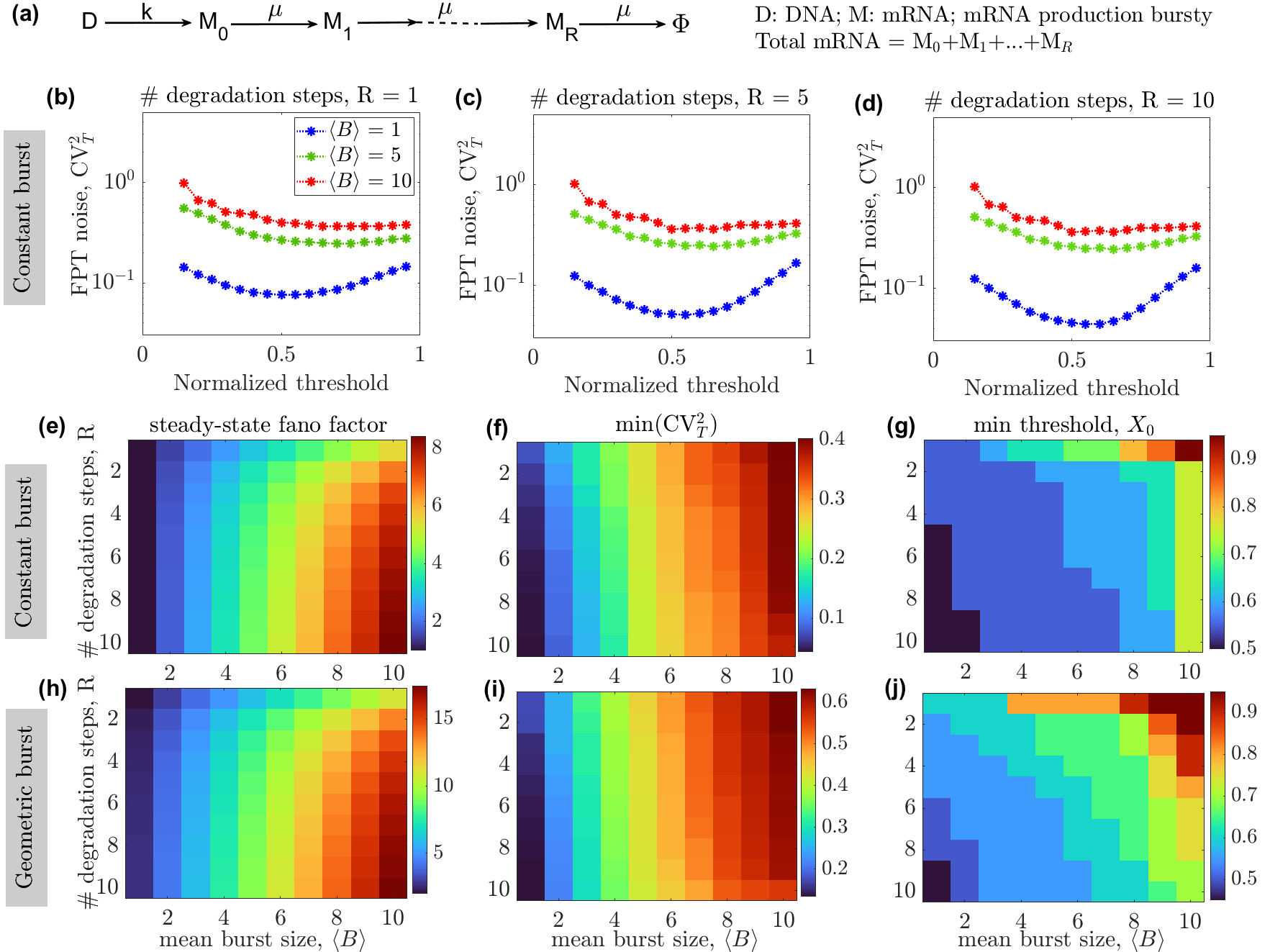}
\caption{\textbf{FPT noise for multi-step mRNA degradation and bursty single-step production.} \textbf{(a)} The schematic of the gene network, where transcription is a single-step process with bursty production of size $B$ and frequency $k$, and degradation is an $ R$-step process with equal rate $\mu$. 
FPT noise with normalized threshold ($X/\langle x_s\rangle$) for different mean burst size, $\langle B\rangle$ when burst is constant and number of degradation steps, \textbf{(a)} $R$ = 1, \textbf{(b)} $R$ = 5, and \textbf{(c)} $R$ = 10.
\textbf{(e)} Steady-state Fano factor of mRNA level, \textbf{(f)} minima value of FPT noise, and \textbf{(g)} minima value of normalized threshold in the $\langle B\rangle-R$ plane, obtained from SSA simulation \cite{gillespie1977exact} over $10^3$ realizations. \textbf{(e-g)}
We consider constant burst, i.e. $\sigma_B^2 = 0$, \textbf{(h-j)} and geometric burst, i.e, $\sigma_B^2 = \langle B\rangle(1+\langle B\rangle)$.
Parameter values: $\mu = R/10$ to keep the mRNA lifetime fixed at 10 and $k = \langle x_s\rangle\mu/(R\langle B\rangle)$ to keep mean steady-state mRNA number fixed at 50 for varying R and $\langle B\rangle$. 
}
\label{fig:R_B_plane}
\end{figure}

\subsection{Exact Variance Dynamics}
\label{app:variance_multistep_decay}

We next derive the exact variance dynamics of the total mRNA abundance.
The lifetime of an individual mRNA is Erlang distributed with shape parameter $R$ and rate $\mu$. Its survival probability at age $u$ is
\begin{equation}
S_R(u)
=
e^{-\mu u}
\sum_{j=0}^{R-1}\frac{(\mu u)^j}{j!}.
\label{app:eq:erlang_survival}
\end{equation}
Consider a burst that occurred at time $u$ before observation. Conditional
on the burst size $B$, the number of mRNAs from this burst that survive
until the observation time, denoted by $Y(u)$, follows a binomial
distribution:
\begin{equation}
Y(u)\mid B
\sim
\operatorname{Binomial}\!\left(B,S_R(u)\right).
\end{equation}
Its first two conditional moments are
\begin{align}
\left\langle Y(u)\mid B\right\rangle
&=
B S_R(u),
\\
\left\langle Y^2(u)\mid B\right\rangle
&=
B S_R(u)\left[1-S_R(u)\right]
+
B^2 S_R^2(u).
\end{align}
Averaging over the burst-size distribution gives
\begin{align}
\left\langle Y(u)\right\rangle
&=
\langle B\rangle S_R(u),
\\
\left\langle Y^2(u)\right\rangle
&=
\langle B\rangle S_R(u)
+
\left(
\langle B^2\rangle-\langle B\rangle
\right)S_R^2(u).
\label{app:eq:surviving_burst_second_moment}
\end{align}
Since the bursts arrive according to a Poisson process with rate $k$,
the variance of the total mRNA abundance is
\begin{equation}
\operatorname{Var}[x(t)]
=
k\int_0^t
\left\langle Y^2(u)\right\rangle du.
\label{app:eq:compound_poisson_variance}
\end{equation}
Using Eq.~\eqref{app:eq:surviving_burst_second_moment}, this becomes
\begin{equation}
\operatorname{Var}[x(t)] = \langle B\rangle k\int_0^t S_R(u)\,du + k\left(\langle B^2\rangle-\langle B\rangle\right) \int_0^t S_R^2(u)\,du.
\label{app:eq:exact_transient_variance}
\end{equation}
We define
\begin{equation}
I_2(t)
=
\int_0^t S_R^2(u)\,du.
\label{app:eq:second_survival_integral}
\end{equation}
Substituting Eq.~\eqref{app:eq:erlang_survival} into Eq.~\eqref{app:eq:second_survival_integral} gives
\begin{align}
I_2(t)
&=
\frac{1}{\mu}
\sum_{i=0}^{R-1}
\sum_{j=0}^{R-1}
\frac{(i+j)!}{i!j!2^{i+j+1}}
\times
\left[
1-e^{-2\mu t}
\sum_{\ell=0}^{i+j}
\frac{(2\mu t)^\ell}{\ell!}
\right].
\label{app:eq:I2_finite_sum}
\end{align}
Equations~\eqref{app:eq:exact_transient_variance} and
\eqref{app:eq:I2_finite_sum} provide the exact variance dynamics for an
arbitrary burst-size distribution with finite first and second moments.

For a constant burst of size $B=b$,
\begin{equation}
\langle B\rangle=b,
\qquad
\langle B^2\rangle=b^2,
\end{equation}
and hence
\begin{equation}
\boxed{
\operatorname{Var}[x(t)]
=
\langle x(t)\rangle
+
k b(b-1)I_2(t).
}
\label{app:eq:variance_constant_burst}
\end{equation}
In particular, for unit bursts, $b=1$,
\begin{equation}
\operatorname{Var}[x(t)]
=
\langle x(t)\rangle,
\qquad
\mathrm{FF}(t)=1,
\label{app:eq:unit_burst_poisson}
\end{equation}
at all times, independently of the number of degradation steps.

Taking the long-time limit of Eq.~\eqref{app:eq:I2_finite_sum} gives
\begin{align}
I_2(\infty)
&=
\frac{1}{\mu}
\sum_{i=0}^{R-1}
\sum_{j=0}^{R-1}
\frac{(i+j)!}{i!j!2^{i+j+1}}
\nonumber\\
&=
\frac{R}{\mu}
\left[
1-\frac{\binom{2R}{R}}{4^R}
\right].
\label{app:eq:I2_steady_state}
\end{align}
Using
\begin{equation}
\langle x_s\rangle
=
\frac{k\langle B\rangle R}{\mu},
\end{equation}
the steady-state variance becomes
\begin{equation}
\operatorname{Var}[x_s]
=
\langle x_s\rangle
\left\{
1+
\frac{
\langle B^2\rangle-\langle B\rangle
}{
\langle B\rangle
}
\left[
1-\frac{\binom{2R}{R}}{4^R}
\right]
\right\}.
\label{app:eq:steady_state_variance_general}
\end{equation}
Therefore, the steady-state Fano factor is
\begin{equation}
\boxed{
\mathrm{FF}_s
=
1+
\frac{
\langle B^2\rangle-\langle B\rangle
}{
\langle B\rangle
}
\left[
1-\frac{\binom{2R}{R}}{4^R}
\right].
}
\label{app:eq:steady_state_fano_general}
\end{equation}
For constant burst of size $b$, this reduces to
\begin{equation}
\boxed{
\mathrm{FF}_s
=
1+
(b-1)
\left[
1-\frac{\binom{2R}{R}}{4^R}
\right].
}
\label{app:eq:steady_state_fano_constant_burst}
\end{equation}

Finally, the exact transient Fano factor is
\begin{equation}
\mathrm{FF}(t)
=
\frac{\operatorname{Var}[x(t)]}{\langle x(t)\rangle}
=
1+
\frac{
k\left(\langle B^2\rangle-\langle B\rangle\right)I_2(t)
}{
\langle x(t)\rangle
},
\qquad t>0.
\label{app:eq:transient_fano}
\end{equation}
These exact abundance moments are used to validate the stochastic
simulations. 

\section{mRNA Mean and Noise Dynamics with Multistep Burst Arrival and Degradation}
\label{app:steady_state_noise_NR}

We derive the transient and steady-state moments of the total mRNA
abundance when both burst arrival and degradation proceed through
multiple stages. Throughout, bursts have a fixed positive integer size
$\langle B\rangle$, so that $\langle B^2\rangle=\langle B\rangle^2$.
Individual transcripts degrade independently, and $x(t)$ denotes the
total number of transcripts across all degradation stages.

\subsection{Model and mean dynamics}

Each burst is generated upon completion of $N$ sequential transcriptional
stages, each with rate $k$. The interburst interval $\tau$ therefore has
the Erlang density
\begin{equation}
    f_N(\tau)
    =\frac{k^N\tau^{N-1}e^{-k\tau}}{(N-1)!},
    \qquad \tau\geq0,
    \qquad \langle\tau\rangle=\frac{N}{k}.
    \label{app:eq:erlang_arrival_NR}
\end{equation}
Thus, the stationary mean burst-arrival rate is
$\lambda=1/\langle\tau\rangle=k/N$.
Here, $\tau$ denotes an interburst interval, whereas
$\langle\tau\rangle$ denotes its mean.

Each transcript passes through $R$ degradation stages, each with rate
$\mu$, before removal. Its survival probability at age $t$ is
\begin{equation}
    S_R(t)=e^{-\mu t}\sum_{j=0}^{R-1}\frac{(\mu t)^j}{j!},
    \qquad
    \int_0^\infty S_R(t)\,dt=\frac{R}{\mu}.
    \label{eq:mrna_survival}
\end{equation}
For the transient calculation, we take $x(0)=0$ and sample the initial
transcriptional phase uniformly over the $N$ stages. The arrival process
is therefore stationary from $t=0$, although the mRNA population is
initially empty. In particular, the first waiting time is the stationary
residual waiting time, rather than a complete Erlang interburst interval
when $N>1$.

A burst of age $u$ contributes $\langle B\rangle S_R(u)$ surviving
transcripts on average. Integrating over burst ages gives mean mRNA number
\begin{align}
    \langle x(t)\rangle
    &=\lambda\langle B\rangle\int_0^t S_R(u)\,du
    \nonumber\\
    &=\frac{k\langle B\rangle}{N\mu}
    \sum_{j=1}^{R}\left[
        1-e^{-\mu t}\sum_{\ell=0}^{j-1}
        \frac{(\mu t)^\ell}{\ell!}
    \right].
    \label{app:eq:mean_transient_NR_sup}
\end{align}
Taking $t\to\infty$ yields the steady-state mean
\begin{equation}
    \langle x_s\rangle
    =\lambda\langle B\rangle\frac{R}{\mu}
    =\frac{k\langle B\rangle R}{N\mu}.
    \label{app:eq:mean_general_NR_sup}
\end{equation}

\subsection{Exact transient noise}

Correlations between distinct burst arrivals are described by the renewal
density
\begin{equation}
    h_N(u)
    =\sum_{n=1}^{\infty}f_N^{*n}(u)
    =\sum_{n=1}^{\infty}
    \frac{k^{nN}u^{nN-1}e^{-ku}}{(nN-1)!},
    \qquad u>0,
    \label{eq:arrival_renewal_density}
\end{equation}
where $f_N^{*n}$ denotes the $n$-fold convolution of $f_N$.
Conditional on a burst at time zero, $h_N(u)\,du$ is the expected
number of bursts in an interval of width $du$ at lag $u$.
Its Laplace transform is
\begin{equation}
    \widetilde h_N(s)
    \equiv\int_0^\infty e^{-su}h_N(u)\,du
    =\frac{\left(k/(k+s)\right)^N}
    {1-\left(k/(k+s)\right)^N},
    \qquad s>0.
    \label{app:eq:renewal_density_NR}
\end{equation}
For a stationary renewal process, the density of ordered pairs of
distinct bursts separated by a positive lag $u$ is $\lambda h_N(u)$.
The excess over the uncorrelated value is therefore
$\lambda[h_N(u)-\lambda]$.

The number of surviving transcripts from a burst of age $u$ is binomial
with parameters $\langle B\rangle$ and $S_R(u)$. Its second factorial
moment is $\langle B\rangle(\langle B\rangle-1)S_R(u)^2$.
Separating pairs of transcripts produced in the same burst from those
produced in distinct bursts gives
\begin{align}
    \langle x(t)[x(t)-1]\rangle
    ={}&\lambda\langle B\rangle(\langle B\rangle-1)
    \int_0^t S_R(u)^2\,du
    \nonumber\\
    &+2\lambda\langle B\rangle^2
    \int_0^t S_R(u)\int_0^u S_R(v)h_N(u-v)\,dv\,du.
    \label{app:eq:factorial_moment_transient_NR}
\end{align}
Using $\operatorname{Var}[x(t)]=\langle x(t)^2\rangle-\langle x(t)\rangle^2$, and defining
\begin{align}
    I_R(t)&=\int_0^t S_R(u)^2\,du,
    \label{eq:survival_squared_integral}\\
    J_{N,R}(t)&=\int_0^t S_R(u)
    \int_0^u S_R(v)[h_N(u-v)-\lambda]\,dv\,du,
    \label{eq:arrival_correlation_integral}
\end{align}
we obtain the exact transient variance
\begin{equation}
    \operatorname{Var}[x(t)]
    =\langle x(t)\rangle
    +\lambda\langle B\rangle(\langle B\rangle-1)I_R(t)
    +2\lambda\langle B\rangle^2J_{N,R}(t).
    \label{eq:variance_combined_dynamics}
\end{equation}
The corresponding squared coefficient of variation is
\begin{equation}
    CV_x^2(t)
    =\frac{
        \langle x(t)\rangle
        +\lambda\langle B\rangle(\langle B\rangle-1)I_R(t)
        +2\lambda\langle B\rangle^2J_{N,R}(t)
    }{\langle x(t)\rangle^2},
    \qquad t>0.
    \label{eq:CV2_combined_dynamics_sup}
\end{equation}
The restriction $t>0$ is required because $\langle x(0)\rangle=0$.
The term involving $I_R(t)$ accounts for transcripts produced in the
same burst, whereas the term involving $J_{N,R}(t)$ accounts for
correlations between distinct bursts.

For Poisson burst arrivals ($N=1$), $h_1(u)=k=\lambda$ and hence
$J_{1,R}(t)=0$. Equation~\eqref{eq:CV2_combined_dynamics_sup} then reduces to
\begin{equation}
    CV_x^2(t)
    =\frac{1}{\langle x(t)\rangle}
    +\frac{\lambda\langle B\rangle(\langle B\rangle-1)}
    {\langle x(t)\rangle^2}
    \int_0^t S_R(u)^2\,du.
    \label{eq:CV2_poisson_arrival_dynamics}
\end{equation}

\subsection{Steady-state limit}

To evaluate the long-time limit, define the survival overlap
\begin{equation}
    C_R(u)=\int_0^\infty S_R(v)S_R(v+u)\,dv.
    \label{app:eq:survival_overlap_NR}
\end{equation}
Changing variables in Eq.~\eqref{eq:arrival_correlation_integral} gives
\begin{equation}
    \lim_{t\to\infty}J_{N,R}(t)
    =\int_0^\infty[h_N(u)-\lambda]C_R(u)\,du.
    \label{app:eq:J_stationary_NR}
\end{equation}
Thus, the exact steady-state variance is
\begin{align}
    \operatorname{Var}[x_s]
    ={}&\langle x_s\rangle
    +\lambda\langle B\rangle(\langle B\rangle-1)
    \int_0^\infty S_R(u)^2\,du
    \nonumber\\
    &+2\lambda\langle B\rangle^2
    \int_0^\infty[h_N(u)-\lambda]C_R(u)\,du.
    \label{app:eq:variance_general_NR}
\end{align}
For Erlang lifetimes,
\begin{equation}
    q_R\equiv
    \frac{\displaystyle\int_0^\infty S_R(u)^2\,du}
    {\displaystyle\int_0^\infty S_R(u)\,du}
    =1-\frac{\binom{2R}{R}}{4^R}.
    \label{app:eq:qR_definition}
\end{equation}
Dividing Eq.~\eqref{app:eq:variance_general_NR} by
Eq.~\eqref{app:eq:mean_general_NR_sup} yields the exact Fano factor
\begin{equation}
    \mathrm{FF}_{N,R}
    \equiv\frac{\operatorname{Var}[x_s]}{\langle x_s\rangle}
    =1+(\langle B\rangle-1)q_R
    +\frac{2\langle B\rangle\mu}{R}
    \int_0^\infty\left[h_N(u)-\frac{k}{N}\right]C_R(u)\,du.
    \label{app:eq:FF_general_NR_integral}
\end{equation}
The steady-state squared coefficient of variation is therefore
\begin{equation}
    CV_{x_s}^2=\frac{\mathrm{FF}_{N,R}}{\langle x_s\rangle}.
    \label{app:eq:CV_general_NR}
\end{equation}

\subsection{Finite-sum representation and limiting cases}

The correlation integral in Eq.~\eqref{app:eq:FF_general_NR_integral}
admits a finite-sum representation. Define
\begin{equation}
    H_N(s)
    \equiv\widetilde h_N(s)-\frac{\lambda}{s}
    =\frac{1}{(1+s/k)^N-1}-\frac{k}{Ns},
    \qquad s>0.
    \label{app:eq:HN_definition}
\end{equation}
Expanding the survival functions in
Eq.~\eqref{app:eq:survival_overlap_NR} gives
\begin{equation}
    C_R(u)=\frac{e^{-\mu u}}{\mu}
    \sum_{p=0}^{R-1}c_{R,p}(\mu u)^p,
    \label{app:eq:CR_finite_expansion}
\end{equation}
with coefficients
\begin{equation}
    c_{R,p}
    =\frac{1}{p!}
    \sum_{i=0}^{R-1}\sum_{\ell=0}^{R-1-p}
    \frac{(i+\ell)!}{i!\,\ell!\,2^{i+\ell+1}}.
    \label{app:eq:cRp_definition}
\end{equation}
Since
\begin{equation}
    \int_0^\infty[h_N(u)-\lambda]e^{-\mu u}u^p\,du
    =(-1)^p\left.\frac{d^pH_N(s)}{ds^p}\right|_{s=\mu},
    \label{app:eq:HN_derivative_identity}
\end{equation}
the exact Fano factor becomes
\begin{equation}
    \mathrm{FF}_{N,R}
    =1+(\langle B\rangle-1)q_R
    +\frac{2\langle B\rangle}{R}
    \sum_{p=0}^{R-1}c_{R,p}(-\mu)^p
    \left.\frac{d^pH_N(s)}{ds^p}\right|_{s=\mu}.
    \label{app:eq:FF_general_NR_finite}
\end{equation}

For single-step burst arrival ($N=1$), $H_1(s)=0$, and
Eq.~\eqref{app:eq:FF_general_NR_finite} reduces to
\begin{equation}
    \mathrm{FF}_{1,R}=1+(\langle B\rangle-1)q_R.
    \label{app:eq:FF_N1_general_R}
\end{equation}
Thus, $\mathrm{FF}_{1,R}=1$ for $\langle B\rangle=1$, whereas it
increases with $R$ for $\langle B\rangle>1$.
For single-step degradation ($R=1$), $q_1=c_{1,0}=1/2$, giving
\begin{equation}
    \mathrm{FF}_{N,1}
    =\frac{1+\langle B\rangle}{2}
    +\langle B\rangle\left[
        \frac{\left(k/(k+\mu)\right)^N}
        {1-\left(k/(k+\mu)\right)^N}
        -\frac{k}{N\mu}
    \right].
    \label{app:eq:FF_general_N_R1}
\end{equation}

\subsection{Large-copy-number approximation}

At fixed $N$, $R$, and $\langle B\rangle$, the large-copy-number limit
corresponds to $\mu/k\to0$, because
\begin{equation}
    \frac{\mu}{k}
    =\frac{\langle B\rangle R}{N\langle x_s\rangle}.
    \label{app:eq:small_parameter_NR}
\end{equation}
The approximation below concerns this limit; increasing the burst size
alone does not establish $\mu/k\ll1$.
Expanding Eq.~\eqref{app:eq:HN_definition} about $s/k=0$ gives
\begin{equation}
    H_N(s)
    =-\frac{N-1}{2N}
    +\frac{N^2-1}{12N}\frac{s}{k}
    +\mathcal{O}\!\left(\frac{s^2}{k^2}\right).
    \label{app:eq:HN_small_s}
\end{equation}
At leading order, only the $p=0$ term in
Eq.~\eqref{app:eq:FF_general_NR_finite} contributes. Moreover,
Eqs.~\eqref{app:eq:CR_finite_expansion} and
\eqref{app:eq:qR_definition} imply
$c_{R,0}=\mu C_R(0)=Rq_R$. Substitution therefore yields
\begin{equation}
    \mathrm{FF}_{N,R}
    =1+\left(\frac{\langle B\rangle}{N}-1\right)q_R
    +\mathcal{O}\!\left(\frac{\mu}{k}\right),
    \label{app:eq:FF_general_NR_large_mean}
\end{equation}
or, equivalently,
\begin{equation}
    CV_{x_s}^2
    \simeq\frac{1}{\langle x_s\rangle}
    \left[1+\left(\frac{\langle B\rangle}{N}-1\right)q_R\right].
    \label{app:eq:CV_general_NR_large_mean}
\end{equation}
The remainder in Eq.~\eqref{app:eq:FF_general_NR_large_mean} is stated
for fixed $N$, $R$, and $\langle B\rangle$; it is not a uniform error
bound when these parameters also increase.

Since
$q_{R+1}-q_R=\binom{2R}{R}/[4^R(2R+2)]>0$, the leading-order change
in the Fano factor at fixed $N$ and $\langle B\rangle$ is
\begin{equation}
    \mathrm{FF}_{N,R+1}-\mathrm{FF}_{N,R}
    \simeq\left(\frac{\langle B\rangle}{N}-1\right)
    (q_{R+1}-q_R),
    \label{app:eq:FF_R_trend}
\end{equation}
provided both parameter sets lie within the approximation regime.
Accordingly, the Fano factor decreases with $R$ when
$\langle B\rangle<N$, is independent of $R$ to leading order when
$\langle B\rangle=N$, and increases with $R$ when
$\langle B\rangle>N$. At fixed $\langle x_s\rangle$, the same trends
apply to $CV_{x_s}^2$.

For $\langle B\rangle=1$ and $N>1$, the approximation predicts that
multistep degradation enhances the noise suppression associated with
regular burst arrival. In contrast, for $N=1$ and
$\langle B\rangle>1$, the interburst correlation term vanishes, and
the contribution from transcripts produced in the same burst causes
the Fano factor to increase with $R$, as follows exactly from
Eq.~\eqref{app:eq:FF_N1_general_R}. These opposite trends reflect the
competition between arrival regularity and the shared production
times of transcripts within each burst.


\end{document}